\documentclass[aps,prl,twocolumn,groupedaddress,showpacs]{revtex4}

\usepackage{graphicx}
\usepackage{dcolumn}
\usepackage{bm}
\usepackage{subfigure}

\begin{document}

\preprint{}

\title{Optical Lattice Induced Light Shifts in an Yb Atomic Clock}

\author{Z. W. Barber}
\altaffiliation{also at University of Colorado, Boulder, CO, 80309}
\email{zbarber@boulder.nist.gov}
\author{J. E. Stalnaker}
\altaffiliation{present address: Department of Physics and Astronomy, Oberlin College, Oberlin OH 44074}
\author{N. D. Lemke}
\altaffiliation{also at University of Colorado, Boulder, CO, 80309}
\author{N. Poli}
\altaffiliation{LENS and Dipartimento di Fisica, Universit\`a di
Firenze, INFN - sezione di Firenze - 50019 Sesto
Fiorentino, Italy}
\author{C. W. Oates}
\author{T. M. Fortier}
\author{S. A. Diddams}
\author{L. Hollberg}
\author{C. W. Hoyt}
\altaffiliation{present address: Department of Physics, Bethel University, St. Paul MN 55112}
\affiliation{National Institute of Standards and Technology\\325
Broadway, Boulder, CO 80305}
\thanks{Official contribution of the National Institute of Standards and Technology of the U.S. Department of Commerce; not subject to copyright.}
\author{A. V. Taichenachev}
\author{V. I. Yudin}
\affiliation{Institute of Laser Physics SB RAS, Novosibirsk 630090, Russia\\ and Novosibirsk State University, Novosibirsk 630090, Russia}
\date{\today}

\begin{abstract}
We present an experimental study of the lattice induced light shifts on the $^1S_0\rightarrow\,^3P_0$ optical clock transition ($\nu_{clock}\approx518$\,THz) in neutral ytterbium.  The ``magic'' frequency, $\nu_{magic}$, for the $^{174}$Yb isotope was determined to be $394\,799\,475(35)$MHz, which leads to a first order light shift uncertainty of 0.38\,Hz on the 518\,THz clock transition.  Also investigated were the hyperpolarizability shifts due to the nearby $6s6p\,^3P_0 \rightarrow$ $6s8p\,^3P_0$, $6s8p\,^3P_2$, and $6s5f\,^3F_2$ two-photon resonances at 759.708\,nm, 754.23\,nm, and 764.95\,nm respectively.  By tuning the lattice frequency over the two-photon resonances and measuring the corresponding clock transition shifts, the hyperpolarizability shift was estimated to be $170(33)$\,mHz for a linear polarized, $50\,\mu$K deep, lattice at the magic wavelength.  In addition, we have confirmed that a circularly polarized lattice eliminates the $J=0\rightarrow J=0$ two-photon resonance.  These results indicate that the differential polarizability and hyperpolarizability frequency shift uncertainties in a Yb lattice clock could be held to well below $10^{-17}$.    
\end{abstract}


\maketitle
The latest generation of atomic clocks are based on narrow optical transitions in atoms or ions.  Moving from microwaves to optical frequencies greatly increases line quality factors, and is enabled by methods of laser cooling and trapping, highly stabilized CW laser sources, and the rapid advances in optical frequency comb technology that allow high accuracy comparison of frequencies ranging from the RF to the UV.  Optical lattice clocks, by confining atoms to the Lamb-Dicke regime, have the potential to achieve very high accuracy (systematic uncertainty less than $10^{-17}$), while at the same time achieving high stability through the use of large numbers of atoms.  

Although most initial lattice clock studies have focused on Sr \cite{Katori03,LeTargat06,Takamoto06,Boyd07}, the alkaline earth-like Yb is also an excellent candidate for an optical lattice clock \cite{Porsev04,Hong05b,Barber06}.  In addition to being relatively simple to laser cool and trap, Yb has a spin-1/2 isotope ($^{171,173}$Yb, $I=1/2,5/2$ respectively) and works well for magnetically-induced spectroscopy of the spin zero isotopes\cite{Taichenachev06}, either of which can be used to reduce experimental complexities due to the nuclear substructure.  In addition to clocks, Yb trapped in optical lattices is of interest for studies of Boson-Fermion interactions\cite{Fukuhara07}, and quantum information\cite{Hayes07}.  In this Letter, we report the first accurate measurement of the magic wavelength for the $^1S_0\rightarrow\,^3P_0$ clock transition of Yb.  We also find that the hyperpolarizability shifts caused by two-photon transitions nearly resonant with the magic wavelength are comparable to those for Sr when scaled in terms of trap depth\cite{Brusch06}.  These measurements are essential to building an accurate optical lattice clock.  The lattice induced light shifts of our Yb clock can currently be determined to $ < 8\times10^{-16}$ in fractional frequency, and there exists no practical limit to reducing this to the $10^{-18}$ level.

A major concern for the accuracy of optical lattice clocks is the ability to cancel the large light shifts created by the trapping potential of the lattice.  Practical optical lattice depths for Yb are $50$ to $500\,E_r$ ($E_r$ is the lattice recoil energy), which means the corresponding light shifts of 100\,kHz to 1\,MHz must be canceled to a high degree. 
The light shift, $\Delta\omega$, on the clock transition due to the lattice can be described by a power series expansion in the electric field strength $E$:
\begin{equation}
\Delta\omega = \alpha\left(\omega_l,\mathbf{e}\right)E^2 + \gamma\left(\omega_l,\mathbf{e}\right)E^4 + \ldots,
\label{Domega}
\end{equation}  
where $\alpha$ and $\gamma$ are the differential polarizability and hyperpolarizability between the ground and excited states respectively, and depend on the lattice frequency, $\omega_l$, and polarization vector $\mathbf{e}$.  The first term dominates, and the design of optical lattice clocks is such to minimize this polarizability term be the choice of polarization insensitive $J=0 \rightarrow J=0$ clock transitions and determining the lattice wavelength where $\alpha$ vanishes.  At this magic wavelength the frequency uncertainty of the first order lattice induced light shift could be held below $10^{-17}$ \cite{Porsev04}.  

\begin{figure}
\includegraphics[width=8cm]{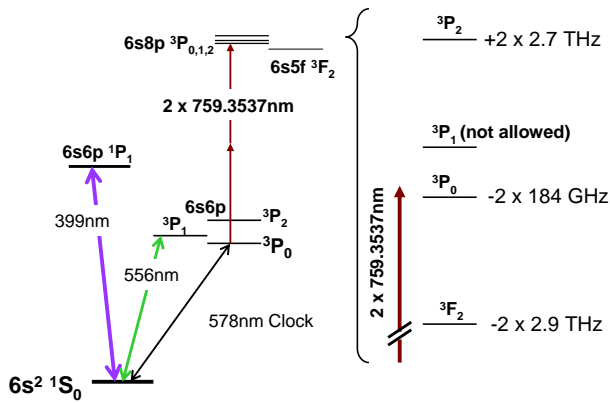}
\caption{Relevant energy levels of Yb, including expanded view of states nearly resonant through two-photon excitation at the magic wavelength.  The two-photon detunings are expressed in terms of the one photon frequency.}
\label{EnergyDiagram}  
\end{figure}  
Higher order light shifts have also been shown theoretically to be below the $10^{-17}$ level, barring near resonant terms in the hyperpolarizability from nearby two-photon transitions \cite{Porsev04}.  However, in Yb the magic wavelength is in the vicinity of three two-photon resonances from the $6s6p\,^3P_0$ excited clock state to higher lying $6s8p\,^3P_j$ and $6s5f\,^3F_2$ states, with the closest a third of a nanometer away (see Fig. \ref{EnergyDiagram}).  Our simple estimate of the strength of this $^3P_0\leftrightarrow\,^3P_0$ transition suggested that the hyperpolarizability shift could have been as large as one hertz at the magic wavelength.  Until now, detailed studies of lattice induced light shifts in Yb have not been performed, and the magic wavelength had been only roughly located to near 759.35\,nm \cite{Barber06}.

Most of our current work utilizes the $^{174}$Yb isotope and the magnetically induced spectroscopic (MIS) method \cite{Taichenachev06,Barber06}. After laser cooling and trapping on both the strong $^1P_1$ singlet line and the weaker $^3P_1$ intercombination line, we load roughly 10\,000 atoms into an optical lattice at temperature of $\approx 15\,\mu$K.  The 1D lattice is formed in the horizontal plane by focusing approximately 1\,W of light to a waist of $\approx30\,\mu$m, then retroreflecting the beam with a curved mirror.  This produces a standing wave lattice consisting of a series of pancake shaped wells with a trap depth of $\sim50\,\mu$K ($500\,E_r$).  After the preparation stage, a static magnetic field ($\sim10\,$mT) is turned on to mix the upper $^3P_0$ clock state with the $^3P_1$ state, which provides a nonzero excitation probability between the two states (MIS method).  While the bias field is on, the atoms are probed along the lattice direction with a frequency stabilized laser at the clock wavelength of 578.42\,nm.  Excitations to the upper clock state are detected by measuring the depletion of the ground state population after the excitation pulse.  With a field of $\approx 1.5$\,mT and about 12\,$\mu$W of probe light focused to a 35\,$\mu$m spot, we generate a $\pi$ pulse in 64\,ms.  For operation as a frequency reference, the stable clock laser is locked to the atomic resonance by use of an acousto-optic modulator to remove the slow drift of the reference cavity.  Further details of the experimental apparatus can be found in reference \cite{Barber07}.  

For the measurements of the magic wavelength, we compared our clock frequency shifts to the Ca clock\cite{Guido06} using an octave spanning Ti:Sapphire optical frequency comb \cite{Fortier06}.  The Ca clock runs in a simplified mode that provides high stability (averaging down to $3\times10^{-16}$ fractional frequency instability in $\approx200$\,s) and high reliability.  In addition, comb was used as a reference to stabilize the frequency of the lattice laser.

\begin{figure}
\includegraphics[width=8cm]{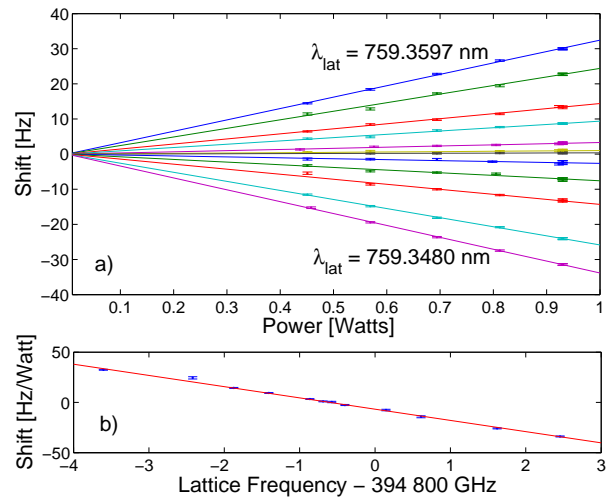}
\caption{ a) The shift of the Yb optical clock frequency versus lattice power for twelve separate lattice wavelengths.  As the lattice wavelength is increased, the differential polarizability (slope of the lines) goes from large and negative, through zero at the magic frequency, $\nu_{magic} = 394\,799\,475(35)$\,MHz, ($\lambda_{magic} \approx 759.3537\,$nm), to large and positive.  The data are individually offset so the fit lines pass through the origin. b) Differential polarizability plotted as a function of lattice frequency.  The fit line was used to interpolate the magic frequency with an uncertainty of 35\,MHz.}
\label{ShiftvInt}
\end{figure}

A rough determination of the magic wavelength was made by observing the width and asymmetry of the spectroscopic lines of the clock transition.  Then the clock frequency shift versus lattice intensity for twelve lattice wavelengths was measured by counting the beat frequency between our Yb locked laser and the comb, which was stabilized to the Ca clock (see Fig.\ref{ShiftvInt}a).  The slope of each wavelength set, which is proportional to the differential polarizability at that wavelength, was extracted with a linear weighted least squares fit.  These slopes were then plotted as a function of the lattice frequency (see Fig. \ref{ShiftvInt}b).  The magic frequency is $\nu_{magic} = 394\,799\,475(35)$\,MHz, or $\lambda_{magic}\approx 759.3537\,$nm, as determined by the zero crossing. Combined with the slope  ($d\alpha/d\nu_{magic}=-22(1)$\,mHz/($E_r\cdot$GHz)), the estimated error gives an uncertainty in the shift of 0.38\,Hz, for a lattice depth of $500\,E_r$.  This uncertainty is limited only by the data presented here.  With additional measurements, the magic wavelength and first order light shift uncertainty will be continually refined and reduced by inter-clock comparisons.  Importantly, a 100\,kHz measurement uncertainty in the magic wavelength is feasible, which would produce an clock shift uncertainty of less than $10^{-18}$.

With the first term in Eq. \ref{Domega} minimized by operating at the magic wavelength, it is important to determine the size of the second term.  The largest contributors to the hyperpolarizability are the near resonant two-photon transitions from the upper clock state to the $6s8p ^3P_{0,2}$ states at 759.7082\,nm ($J=0$) and 754.23\,nm ($J=2$) and to the $6s5f ^3F_2$ state at 764.95\,nm.  (The two-photon transition to the $6s8p ^3P_1$ state at 758.41\,nm is not allowed, due to selection rules.)  The closest two-photon transition is the $^3P_0 \rightarrow\,^3P_0$ transition, which is -184\,GHz (-368\,GHz two-photon detuning) from the magic wavelength (see Fig. \ref{EnergyDiagram}).  

To measure the effect of the two-photon transition on the clock frequency, we tuned the lattice close to this resonance and measured the $^1S_0\rightarrow\,^3P_0$ clock shift, similar to the approach used for Sr \cite{Brusch06}. 
With the lattice detuned from the magic wavelength, the spectroscopic lines observed on the clock transition were slightly asymmetric and broadened to about 1.5\,kHz.  For these broad lines our reference cavity was sufficiently stable (400\,Hz linear drift over a few hours), so that the Ca clock was not needed as a reference.

\begin{figure}
\includegraphics[width=8cm]{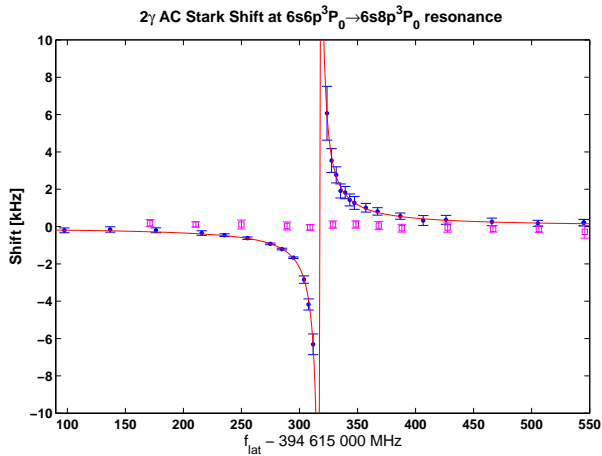}
\caption{AC Stark shift on the clock transition with the lattice tuned near the $6s6p ^3P_0 \leftrightarrow 6s8p ^3P_0$ two-photon resonance with linear (solid points) and circular (open squares) lattice polarization.  Extrapolating the dispersion shaped fit to the magic wavelength gives a shift of $200(30)$\,mHz for a $500\,E_r$ trap depth. The resonance disappears in the circular polarized case due to selection rules.}
\label{2p3p0}
\end{figure}

The light shift on the clock transition with the lattice frequency tuned close to the $6s6p\,^3P_0\rightarrow6s8p\,^3P_0$ two-photon resonance is shown in Fig. \ref{2p3p0}.  At each point of lattice laser frequency, we scanned over the clock resonance 6 to 8 times.  The line shapes were then fit with a Lorentzian line shape to determine the line center.  The measured clock frequency shifts versus lattice wavelength are then fit with a dispersion line shape, which when extrapolated to the measured magic wavelength gives a clock shift of $0.80(12)\,\mu$Hz/$E_r^2$.  For our lattice depth of $\approx500\,E_r$, this leads to a clock shift of $200(30)$\,mHz or $4\times10^{-16}$ fractionally.  It must be noted that the associated uncertainty is mainly due to our determination of the peak lattice intensity, as the extrapolated fit supports a much smaller uncertainty.  

We also performed a similar set of light shift measurements with circularly polarized light.  In this case, the $^3P_0\rightarrow\,^3P_0$ two-photon transition vanishes because of two-photon selection rules.  Although the circular polarization was imperfect ($\sim 10\,\%$ linear component from birefringence of the vacuum windows), the resonance was not discernible in either the width or centers of the lines (see Fig. \ref{2p3p0}).  With the use of an even isotope, the first order light shift is very insensitive to the polarization, making the lattice ellipticity a free parameter that could be used to minimize or cancel the hyperpolarizability shift\cite{Taichenachev06b,Yudin07}.

\begin{figure}
\includegraphics[width=8cm]{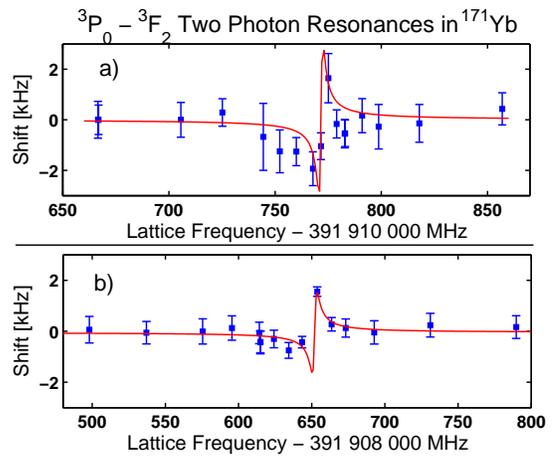}
\caption{AC stark shifts on the Yb optical clock frequency with the lattice laser tuned near the $6s6p^3P_0\rightarrow6s5f^3F_2$ two-photon resonances.  The two resonances are the $F=5/2,3/2$ hyperfine components of the $^3F_2$ state in the $^{171}$Yb isotope, and are split by 4240(20)\,MHz.  The fit gives a total clock shift of 3.5\,mHz when extrapolated to the magic wavelength.}
\label{3p03f2}
\end{figure}

In comparison to the $^3P_0$ two-photon resonance the $^3P_2$ and $^3F_2$ two-photon resonances are much further detuned, which would suggest their associated hyperpolarizability shifts should be significantly smaller.  To verify this, we switched to spectroscopy of the $^{171}$Yb isotope because large light shift broadening due with the lattice tuned to 754.23\,nm and 764.95\,nm made efficient excitation of the clock transition with the MIS technique difficult.  For the $^{171}$Yb isotope, much larger Rabi frequencies could be generated resulting in higher contrast spectroscopic lines.

The resulting light shift measurements of the clock resonance near the $^3F_2$ two-photon resonance are shown in Fig. \ref{3p03f2}.  The separate two-photon resonances are the two hyperfine components caused by the $I=1/2$ nuclear spin of $^{171}$Yb, which are split by 4240(20)\,MHz (2120(10)\,MHz for the lattice frequency).  The large error bars of the individual points in Fig.\,\ref{3p03f2} are caused by the large linewidths and asymmetries of the observed lines.  By fitting the clock shift versus lattice frequency for the two two-photon resonances and extrapolating to the magic wavelength, the total hyperpolarizability shift from the $^3F_2$ state is 3.5(3.5)\,mHz for typical trap depths.  However, because the fits are dominated by only a few points, we place a conservative upper limit of $10\,$mHz on the total shift for our $500\,E_r$ deep lattice.

The two hyperfine components of the $^3P_0\rightarrow\,^3P_2$ two-photon resonance were located at lattice frequencies of 380\,480\,706\,MHz and 380\,480\,602\,MHz.  Clock shift measurements of these resonances (not shown) give an associated hyperpolarizability shift at the magic wavelength of -38(8)\,mHz for a 500\,$E_r$ trap.  This is consistent with a calculation of the shift using knowledge of the $^3P_0$ resonance shifts and angular momentum considerations.  

Totaling 170(33)\,mHz for a 500\,$E_r$ lattice, the combined magnitude of hyperpolarizability shifts is non-negligible but manageable. Indeed, the size of the shifts are comparable to those of Sr when scaled in terms of $E_r$.  The shifts are more significant in our case than for the present Sr lattice clocks because we perform spectroscopy with considerably higher lattice intensities.  However, we anticipate using significantly shallower traps in the future, which with the intensity squared dependence of the hyperpolarizability shift will greatly reduce the magnitude of the shifts.  Lattice depths of $50\,E_r$, or $\approx5\,\mu$K, should be feasible for Yb assuming that laser cooling near the Doppler limit ($\sim 4\mu$K) is achieved.  In addition, the use of vertically oriented or 3D lattices could be of benefit at lower trap depths.  At $50\,E_r$ the total hyperpolarizability shift would be less than $5\times10^{-18}$ in fractional frequency and with uncertainty less than $10^{-18}$.  Thus we do not anticipate the need for more precise measurements of the hyperpolarizability coefficients unless we choose to keep the trap depths above $200\,E_r$.  These hyperpolarizability shift estimates should apply to any of the isotopes of Yb, as the two-photon detunings are much larger than the isotope shifts or the hyperfine splittings.  Finally, the relative size of these two-photon shifts allows for a magic ellipticity where the differential hyperpolarizability coefficient, $\gamma$, is zero\cite{Taichenachev06b}; however, good control of the polarization would be required.  
 
In this Letter, we report the reduction of the total lattice-induced shift uncertainty to a fractional frequency of less than $8\times10^{-16}$.  The magic wavelength has been determined with an uncertainty of 35\,MHz, which for our largest trap depths (500\,$E_r$) produces an uncertainty of $7\times10^{-16}$ fractional frequency, and further measurements will reduce this uncertainty.  The outstanding issue of the possibly large hyperpolarizability due to nearby two-photon resonances has also been resolved.  The hyperpolarizability shift uncertainty has been measured to less than $7\times10^{-17}$ of the clock frequency.  We see no obstacle to reducing the lattice shift uncertainties of an Yb optical lattice clock to well below $10^{-17}$.

The authors thank Kristin Beck for her contributions to these measurements. A.V.T. and V.I.Yu. are supported by
INTAS-SBRAS (06-1000013-9427), RFBR (07-02-01230, 07-02-01028, 08-02-01347, 08-02-01108), and by Presidium SB RAS.


\end{document}